\documentclass[a4paper,12pt]{article}
\usepackage{graphicx}
\usepackage{graphics}
\usepackage{epsf}
\begin{document}
\newcommand{\np}{Nucl.\,Phys.\,}
\newcommand{\pl}{Phys.\,Lett.\,}
\newcommand{\pr}{Phys.\,Rev.\,}
\newcommand{\prl}{Phys.\,Rev.\,Lett.\,}
\newcommand{\prep}{Phys.\,Rep.\,}
\newcommand{\zp}{Z.\,Phys.\,}
\newcommand{\sovjnp}{{\em Sov.\ J.\ Nucl.\ Phys.\ }}
\newcommand{\nuclinst}{{\em Nucl.\ Instrum.\ Meth.\ }}
\newcommand{\annp}{{\em Ann.\ Phys.\ }}
\newcommand{\intjmp}{{\em Int.\ J.\ of Mod.\  Phys.\ }}
\newcommand{\ra}{\rightarrow}
\def\micro{{\tt micrOMEGAs}}
\def\darksusy{{\tt DarkSusy}}
\def\suspect{{\tt SuSpect}}
\def\softsusy{{\tt SoftSusy}}
\def\comphep{{\tt CompHEP}}
\def\isasugra{{\tt ISASUGRA/Isajet}}
\def\isajet{{\tt Isajet}}
\def\neuto{\tilde{\chi}_1^0}
\def\mneuto{m_{\tilde{\chi}_1^0}}
\def\bsgamma{b\rightarrow s\gamma}

\begin{center}
{\large {\bf micrOMEGAs: recent developments}}

\begin{tabular}[t]{c}

{\bf G.~B\'elanger$^{1}$, F.~Boudjema$^{1}$,  A. Pukhov$^{2}$,
A. Semenov$^{1}$}
 \\
{\it 1. Laboratoire de Physique Th\'eorique}
{\large LAPTH}
\footnote{URA 14-36 du CNRS, associ\'ee  \`a
l'Universit\'e de Savoie.}\\
 {\it Chemin de Bellevue, B.P. 110, F-74941 Annecy-le-Vieux,
Cedex, France.}\\

{\it 2. Skobeltsyn Institute of Nuclear Physics, 
Moscow State University} \\ {\it Moscow 119992,
Russia }\\

\end{tabular}
\end{center}


\vspace{-.5cm}
\begin{abstract}
\baselineskip=14pt
The program \micro~ that  calculates the relic density of the lightest supersymmetric
 particle (LSP) in the MSSM is presented. The impact of 
 coannihilation channels and of
 higher order corrections to Higgs widths is stressed.
 The  dependence on the RGE   
 code used    to calculate the soft parameters is also discussed.
 \end{abstract}
\
\vspace{-1cm}

\baselineskip=14pt

\section{Introduction}
The measurements of the relic density of cold dark matter (CDM) have  provided  
 stringent constraints on the
parameters of the R-parity conserving
mininal supersymmetric standard model (MSSM). Hence
a large effort has been devoted to the calculation of the relic density 
in the MSSM, and many private and public programs have been developed
\cite{neutdriver}$^-$\cite{baer}.
Here we present \micro, a program that is publicly available \cite{cpc}.
To make predictions for the relic density
of CDM at the few percent level, special care must be taken to treat
carefully the case where annihilation via a s-channel
resonance can occur as well as the
case of coannihilations where the LSP interacts 
with slightly heavier sparticles.


The  calculation of the relic density  
necessitates the evaluation of a thermally averaged
cross-section. The proper relativistic formalism for
treating this was introduced  in Ref.~8 and 
 proved to be essential  when annihilation through
s-channel pole is important.
The  generalization of this formalism to the case of coannihilations
\cite{EdsjoGondolo}, was implemented in the codes of 
Refs.~2-3, 
for the case of gaugino coannihilations.
We follow basically this formalism.

Coannihilation processes where the LSP interacts 
with slightly heavier sparticles  can occur in principle with any 
supersymmetric particle \cite{Griest}, although in SUGRA models, the most
common coannihilations are with gauginos \cite{Yamaguchi,EdsjoGondolo},
  right-handed sleptons \cite{Ellis-coann,GLP}  
or stops \cite{abdelstop}. 
In micrOMEGAs we include {\em ALL} coannihilation channels, 
in all more than 2800 processes not counting charged conjugate processes.
The tree-level cross-sections
are calculated exactly including the full set of diagrams contributing
to each process.
The calculations of the  cross-sections are based on \comphep \cite{comphep}.
Furthermore we include also some higher order effects, namely
the two-loop corrections to the Higgs mass \cite{FeynHiggs} and the one-loop
QCD corrections to the Higgs width \cite{hdecay}. The latter turns out
 to be particularly important in the large $tan\beta$ region with the enhanced coupling of the Higgs to b quarks.

After the important equations for the calculation of the relic density are summarized,
we give a short description  of the package.
We present  some results and comparisons with 
other programs emphasizing the role of coannihilations 
and Higgs poles. Finally we discuss the impact of the choice of the RGE
code in SUGRA models on the relic density.

\section{Calculation of the relic density}

The calculation of the relic density at present 
necessitates solving the evolution 
equation for the relic abundance, $Y$ 
\begin{equation}
   \frac{dY}{dT}= \sqrt{\frac{\pi  g_*(T) }{45G}} <\sigma v>(Y^2-Y_{eq}^2)
    \label{dydt}
\end{equation}
where $Y_{eq}$ represents the thermal equilibrium abundance.
This equation 
  depends on  $<\sigma v>$, the relativistic thermally averaged
annihilation cross-section, which involves a sum over 
 $\sigma_{ij}$ , the total cross section for annihilation of
a pair of supersymmetric particles into Standard Model
particles.
\begin{equation}
       <\sigma v>=  \frac{ \sum\limits_{i,j}g_i g_j  \int\limits_{(m_i+m_j)^2} ds\sqrt{s}
K_1(\sqrt{s}/T) p_{ij}^2\sigma_{ij}(s)}
                         {2T\big(\sum\limits_i g_i m_i^2 K_2(m_i/T)\big)^2 }\;,
\label{sigmav}
\end{equation}
 The total number of processes involving two 
SUSY particles  into two SM particles exceeds 2800.
In practice,  processes involving the heavier SUSY particles contribute only 
when there is a near mass degeneracy with the LSP
due to  a strong Boltzmann suppression factor.
To speed up the program a given  subprocess  is removed from the sum
(\ref{sigmav}) if the total mass
of the incoming particles is below 
a value 
 defined by the user, typically $\approx 2.5\mneuto$. 

Rather than solving 
for $Y$ numerically, which is extremely
time consuming especially  when we include a great number of processes, 
we  follow the usual procedure of defining 
a freeze-out temperature $T_f$ \cite{GondoloGelmini}.
This approach differs from the one in \darksusy \cite{cpc}. 

\section{Description of  \micro}
\micro~ is a C program that also calls some external FORTRAN functions.
\micro~ relies on {\tt CompHEP} \cite{comphep} for the definition of the
 parameters and the evaluation of all cross-sections.
Only a  small fraction of the available processes are needed for a given model,
 those with a sparticle  close in mass to the LSP. 
To restrict the size of the program,  we include in  our package the program
\comphep \cite{comphep}  which generates, while running, the  
subprocesses needed for a given set of  MSSM parameters.
The generated code  is linked during the run to the main
program and executed. 

The model file used by CompHEP is obtained via LanHEP \cite{lanhep}, a program that
generates the complete set of particles and vertices once given a  Lagrangian.
 In the model used,  the Higgs masses are calculated at
two-loop with {\tt FeynHiggsFast} \cite{FeynHiggs}. 
Higher order QCD corrections to the Higgs widths are  incorporated
by extracting, from HDECAY \cite{hdecay}, 
 effective quark masses $m_q(m_H)$ which are then included in the $Hq\overline{q}$ vertices.

In \micro, there are two options for the input parameters: either the soft parameters
of the supersymmetric Lagrangian at the weak scale or 
the parameters of a SUGRA-type model at the GUT scale. In 
Version 1.1.1, the latter option  is made available  via a link to \isajet.
However, the next upgrade will include links to other codes,
namely \suspect~ and \softsusy, as will be discussed in the last section. 

After  the calculation of the relic density 
is performed, the list of channels that give
the most significant contribution to $\Omega h^2$ are given. 
We also provide subroutines that calculate
various constraints on the MSSM parameters: direct limits from
colliders,  $b\to  s\gamma$ and $(g-2)_\mu$.

\section{Results and Comparisons}

The \micro~ code was extensively  tested against another public package 
for   calculating  the  relic density, {\tt DarkSUSY}.  
The two codes  differ somewhat in the numerical method used for 
solving the density equations, in the number of channels included
(all subprocesses in \micro~) and in the use of
loop-corrected Higgs widths {\footnote{
Sfermions coannihilations will be included 
in upgrades of \darksusy \cite{edsjo_talk}.}}. 
Whenever the coannihilation channels with sfermions and the Higgs pole are
not important we find good agreement with {\tt DarkSUSY}
{\footnote{Complete agreement between  \micro~ and  an improved version of 
\darksusy~
including slepton coannihilation channels was found recently
in Ref.~\cite{mario}.}}.
However, when non gaugino coannihilations are important, we find that the
impact of the extra channels
can be as large as two order of magnitude and depends
critically on the mass difference with the lightest neutralino.
Due to the large cross-sections in
channels involving strongly interacting particles, 
the effect of coannihilation is particularly striking when 
the NLSP is a squark (Fig.~1a).

\begin{figure}[htb]
\vspace{-1.2cm}
\epsfxsize=14.5cm   
\epsfbox{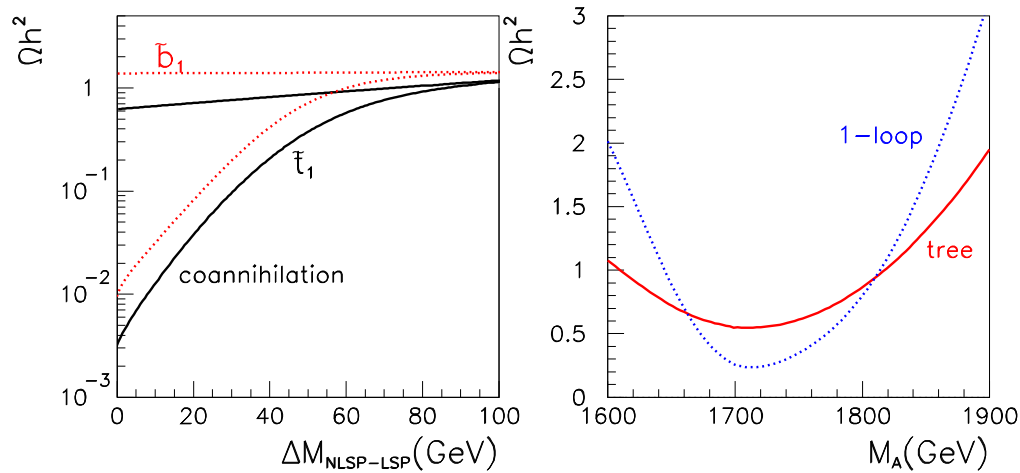}
\vspace{-1.cm}   
\caption{ a)$\Omega h^2$ vs the NLSP-LSP mass difference  
with  $m_{\tilde{u_R}}(m_{\tilde{d_R}})$ as  a free parameter including
coannihilation(full) and without coannihilation(dash). 
The NLSP is the
 $\tilde{t}_1$ ($\tilde{b}_1$). The parameters are the ones of
 Model F;
b)Comparison of tree-level/one-loop treatment of Higgs width.
$\Omega h^2$ vs $m_A$, $\tan\beta=45$, other parameters are the ones of
 Model E.}
\end{figure}


 Near a heavy Higgs
 resonance, we also observe differences with \darksusy, these 
 disappear if we switch to the tree-level width
 option.
The effect of the Higgs width is particularly important at large $\tan\beta$
with the enhanced contribution of the $b$-quark coupling to the heavy scalar
Higgs. One-loop  QCD corrections  which reduce the Higgs width especially at large values of $m_H$ can change $\Omega h^2$  by as much as a
factor 2 (Fig.~1b). 

In the case of SUGRA models, we find qualitative agreement
with Ref.~\cite{benchmark} particularly for small values of $\tan\beta$.
Significant differences are found at large $\tan\beta$.  This
can be due to the RGE code used to calculate the soft parameters
at the weak scale, 
as  will be discussed next.

\section{RGE code and relic density}

To determine the influence 
 on the RGE code used to determine the soft MSSM
parameters on the calculation of the relic density,
  we have compared three of
the most widely used RGE codes: \isajet, \suspect~ and \softsusy.
The parameters that we expect might influence significantly the value
of the relic density are $\mu$ which determines the gaugino fraction, 
$m_A$ which is 
relevant when neutralino annihilation occurs near the Higgs s-channel pole
and finally the NLSP-LSP mass difference.

\begin{figure}[htb]
\vspace{-1.2cm}
\epsfxsize=14.5cm   
\epsfbox{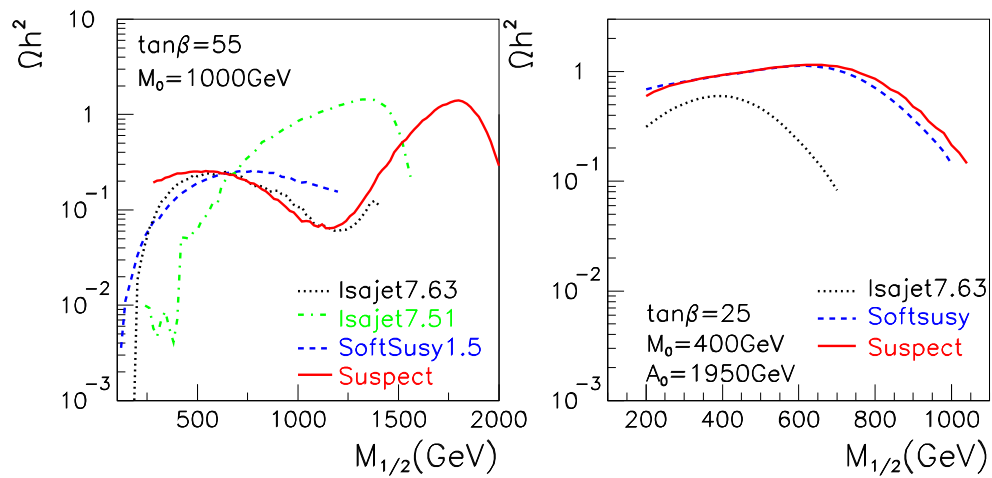}
\vspace{-1.2cm}
\caption{Comparison of \isajet, \softsusy~ and \suspect,
$\Omega h^2$ vs $m_{1/2}$  in a SUGRA model with $\mu>0$ and $m_{top}=175$GeV
a) $\tan\beta=55$ b) $\tan\beta=25, A_0=1950$GeV.
\label{rge}}
\vspace{-.6cm}
\end{figure}

In the notoriously difficult
large $\tan\beta$ region, significant differences in the value of
$m_A$ \cite{Ben_rgecodes} can lead up to 
 factors of two differences in the relic
density (Fig.~\ref{rge}).
At large $M_0$, the relic density calculated in \isajet~ can be 
two orders of magnitude below that in the other two codes due to 
a  significantly lower value for the  parameter $\mu$ \cite{Ben_rgecodes}
(leading to a large  increase in the $\neuto\neuto\ra W^+W^-$ cross-section).
Finally in the coannihilation regions, the relic density can vary by 
an order of magniture even though the differences
between the codes  can be rather small (a few Gev's). 
This occurs at large $M_{1/2}$ and/or large $A_0$ 
where  the $\tilde\tau$ is the NLSP (Fig.~\ref{rge})
\cite{rge_nous}.
In general rather good agreement is found between \softsusy~ and \suspect, whereas
larger differences can be observed with \isajet.

\section{Conclusion}
The package \micro~ that allows to calculate the relic density of the LSP in the MSSM 
is the first program that includes all possible coannihilation
channels.
 Loop corrections to the masses and widths of Higgs particles  are implemented. 
Good agreement with existing calculations is found when  identical  channels
are included and higher order corrections are removed.
 
 The next upgrade of \micro~ 
 will  include more links to programs for calculating RGE, 
 (\softsusy~ and \suspect) as well as new improved routine  to $\bsgamma$.
 Progress has also been made towards including new models beyond the MSSM,
  such as the nMSSM, and additonal modules for calculating constraints
  on the MSSM such  as $B_s\ra \mu^+ \mu^-$ will be available.

\section*{Acknowledgments}
This work was supported in part by the PICS-397.

\end{document}